\documentclass[a4paper,11pt]{article}
\usepackage{jheppub}

\usepackage[utf8]{inputenc}
\usepackage{physics}
\usepackage{amssymb}
\usepackage{slashed}
\usepackage{graphicx}

\title{\boldmath%
Strange-quark mass effects in the $B_s$ meson's light-cone distribution amplitude
}

\author{Thorsten Feldmann,}
\emailAdd{thorsten.feldmann@uni-siegen.de}
\author{Philip L\"ughausen}
\emailAdd{lueghausen@physik.uni-siegen.de}
\author{and Nicolas Seitz}
\emailAdd{nicolas.seitz@uni-siegen.de}

\affiliation{%
Theoretische Physik 1,
Universität Siegen,\\
Walter-Flex-Straße 3,
D-57068 Siegen, Germany
}

\abstract{%
We investigate the differences between the light-cone distribution amplitudes (LCDAs) of $B_s$ mesons and $B_q$ mesons (with $q=u,d$) induced by a non-vanishing strange-quark mass $m_s\neq 0$ (compared to $m_q \simeq 0$). 
To this end we consider the so-called ``radiative tail'' which is related to the 
short-distance expansion of the relevant light-ray operators in heavy-quark effective theory. We extend the 
calculation of the according matching coefficients, including operators 
linear in $m_s$ for both the leading and sub-leading 2-particle LCDAs.
Based on a generic parameterization for the leading LCDA,
we discuss the effect on its shape on a quantitative level, 
and compare our findings with recent results on the inverse moments of the $B_q$ and $B_s$ LCDAs from QCD sum rules.
}

\begin{document}

\preprint{SI-HEP-2023-12, P3H-23-038}

\maketitle

\section{Introduction}

The light-cone distribution amplitudes (LCDAs) of the $B_s$ meson
enter as fundamental hadronic input functions for exclusive $B_s$-meson decays into light energetic particles in the QCD factorization approach. Important examples are non-leptonic 2-body decays, see e.g.\ Refs.~\cite{Beneke:2003zv,Williamson:2006hb,Cheng:2009mu,Huber:2021cgk}, and rare radiative and semileptonic decays, see e.g.\ Refs.~\cite{Ali:2007sj,Beneke:2020fot}.
Comprehensive reviews about the phenomenology of these decays and more references can be found e.g.\ in Refs.~\cite{LHCb:2012myk,Belle-II:2018jsg}.

In the past, the LCDAs for $B_u^\pm$ and $B_d^0$ mesons have been studied intensively, which has led to a number of phenomenological and theoretical constraints. Most importantly, the radiative leptonic decay $B^-\to \gamma \ell^- \bar \nu$ at large recoil energy is sensitive to the inverse moment of the leading $B$-meson LCDA, denoted as $\phi_B^+(\omega)$ in the following, where $\omega$ denotes the light-cone projection of the momentum of the 
light quark in the heavy meson.
While an analogous observable does not exist for $B_s$ mesons, the effect of a non-vanishing strange-quark mass on the inverse moment has recently been estimated from a QCD sum-rule analysis \cite{Khodjamirian:2020hob}. This non-perturbative information essentially constrains the difference between the $B_s$ and $B_q$ LCDAs at small and intermediate values of the light-cone momentum $\omega$.
On the other hand, at large values $\omega \gg\Lambda_{\rm QCD}$ the so-called ``radiative tail'' of the LCDAs can be computed in fixed-order perturbation theory. The 1-loop result for the $B_q$ meson has been derived from the analysis of the cut-off dependence of the positive moments in momentum space \cite{Lee:2005gza} and from a short-distance operator-product expansion in position space \cite{Kawamura:2008vq}.
For phenomenological studies in general, and in order to interpolate between the behavior of the LCDAs at small and large values of $\omega$, in particular, one needs explicit parameterizations of the $B_q$ and $B_s$-meson's LCDAs. A generic parameterization based on a systematic expansion in terms of associated Laguerre polynomials has recently been proposed in Ref.~\cite{Feldmann:2022uok}. The expansion coefficients in this parameterization fulfill integral bounds that control the systematic uncertainties from the truncation of the expansion.%
\footnote{%
This parameterization has already been used \cite{Feldmann:2022ixt} in the context of QED corrections to $B_s \to \mu^+\mu^-$ decays, when internal photons resolve the hadronic structure of the $B_s$ meson \cite{Beneke:2017vpq}.
}

In this letter, we derive the strange-quark mass effects on the radiative tail by including the linear terms in $m_s$ in the short-distance expansion of the relevant light-ray operator. The result can be inferred in a straightforward manner from the perturbative calculation of the LCDA for the $B_c$ meson, which has been derived in a non-relativistic setup in Ref.~\cite{Bell:2008er}.
We compute the constraints on the expansion coefficients in the parameterization \cite{Feldmann:2022uok} from the radiative tail, and compare the resulting estimate for the inverse moments of the $B_s$ and $B_q$-meson LCDAs with the expectations from QCD sum rules. 
The outline of the article is as follows. In the next section, we provide the basic definitions of the $B$-meson 2-particle LCDAs, together 
with the main features of the parameterization we are going to use. We also introduce the short-distance expansion of the relevant light-ray operator, 
including the known results for the matching coefficients in the case of vanishing spectator mass.
Section~\ref{sec:match} is devoted to the derivation of the matching coefficient for the dimension-4 operator proportional to the strange-quark mass.
By a detailed comparison we show how the result can be inferred from the calculation of LCDAs in a non-relativistic setup \cite{Bell:2008er}.
In particular, we show how the matching coefficients in position space can easily be obtained from momentum-space Feynman integrals
in dimensional regularization, 
by performing simple subtractions on the level of the Fourier transformation prior to the expansion in $D-4$ dimensions.
In section~\ref{sec:num} we work out the constraints from the short-distance expansion on the LCDAs, using the generic parameterization
of the $B$-meson LCDA advocated for in Ref.~\cite{Feldmann:2022uok}. 
Truncating the expansion after three terms, 
we give numerical results for the expansion coefficients and the resulting inverse moments of the LCDAs 
for the massive and massless case and compare with recent results from QCD sum rules \cite{Khodjamirian:2020hob}. 
We end the article with our conclusions. In Appendix~\ref{sec:radiative-tail} we provide a generic derivation of the short-distance expansion of
a 2-particle light-ray operator with arbitrary Dirac structure from which we can also read off the matching coefficients for 
the subleading 2-particle LCDA of the $B$ meson.
In Appendix~\ref{sec:Bc} we briefly discuss the extrapolation of our results to the case of LCDAs for $B_c$-mesons. In Appendix~\ref{sec:app-dim-5} we study the impact of the 
dimension-5 operators on the LCDA parameterization.

\section{Preliminaries}

\subsection{Definition of 2-particle LCDAs}

The LCDA that appears at leading power in QCD factorization theorems for exclusive $B$-meson decays can be defined as the hadronic matrix element 
of a 2-particle light-ray operator in heavy-quark effective theory (HQET)
\cite{Grozin:1996pq},
\begin{eqnarray}
 i m_B f_B^{\rm HQET}(\mu) \, 
  \phi_B^+(\omega;\mu) &=& \int \frac{d\tau}{2\pi} \, e^{i  \omega  \tau } \, \langle 0| \bar q( \tau n) \, [\tau n,0] \, \slashed n \gamma_5 \, h_v(0)|\bar{B}(v)\rangle  \,,
  \label{Bdef}
\end{eqnarray}
where $v^\mu$ with $v^2=1$ is the four-velocity of a heavy meson, $B=B_q,B_s$, and $f_B^{\rm HQET}(\mu)$ is the decay constant in the static limit. 
Furthermore, $n^\mu$ with $n^2=0$ is a light-like Lorentz vector, and 
$\omega= n \cdot k$ can be viewed as 
the light-cone projection of the 
light spectator-quark momentum $k^\mu$.
Here, for simplicity, we are considering a frame where $v \cdot n =1$.
The support properties of the matrix element 
in (\ref{Bdef}) are   
\begin{eqnarray}
  \phi_B^+(\omega,\mu) \neq 0 \quad & \mbox{for} & \omega \in \left[0,\infty\right) \,,
\end{eqnarray}
An analogous definition for the sub-leading 2-particle LCDA reads
\begin{eqnarray}
	i m_B f_B^{\rm HQET}(\mu) \, 
	\phi_B^-(\omega;\mu) &=& \int \frac{d\tau}{2\pi} \, e^{i  \omega  \tau } \, \langle 0| \bar q( \tau n) \, [\tau n,0] \, \slashed{\bar n} \gamma_5 \, h_v(0)|\bar{B}(v)\rangle  \,,
	\label{Bmdef}
\end{eqnarray}
where $\bar n^\mu= 2 v^\mu - n^\mu$. In the following, we concentrate on the leading LCDA $\phi_B^+(\omega)$. 
In Appendix~\ref{sec:radiative-tail} we repeat the calculation in a generalised way and give analogous results for $\phi_B^-(\omega)$.

\subsection{A generic parameterization for \texorpdfstring{$\phi_B^+(\omega)$}{phi+}}

In this letter we use the following generic parameterization for the leading $B$-meson LCDA \cite{Feldmann:2022uok},
\begin{eqnarray}
	\phi_B^+(\omega,\mu_0) &=& 
	\frac{\omega \, e^{-\omega/\omega_0}}{\omega_0^2} 
	\, \sum_{k=0}^K \frac{a_k(\mu_0)}{1+k} \, L_k^{(1)}(2\omega/\omega_0) \,,
 \label{eq:generic}
\end{eqnarray}
at a low renormalization scale $\mu_0 \sim 1$~GeV.
Here $L_k^{(1)}(z)$ are associated Laguerre polynomials, and $\omega_0 = {\cal O}(\Lambda_{\rm QCD})$ is an auxiliary reference momentum. The expansion coefficients $a_k$ fulfill the integral bound,
\begin{eqnarray}
	\sum_{k=0}^\infty |a_k(\mu_0)|^2 &=& 
	2\omega_0 \, \int_0^\infty d\omega 
	\left( \left|\phi_B^+(\omega,\mu_0) \right|^2 
	+ \omega_0^2 \left| \frac{d\phi_B^+(\omega,\mu_0)}{d\omega} \right|^2 \right) < \infty \,.
 \label{eq:bound}
\end{eqnarray}
Note that the coefficients $a_k$ are dimensionless quantities that implicitly depend on 
$\omega_0$.
In particular, the expansion of the inverse moment in this parameterization reads
\begin{eqnarray}
	\lambda_B^{-1} 
    \equiv \int_0^\infty\frac{d\omega}{\omega} \, \phi_B^+(\omega,\mu_0) &=& 
	\frac{1}{\omega_0} \, \sum_{k=0}^K \frac{1+(-1)^{k}}{2} \, 
	\frac{a_k(\mu_0)}{1+k} = \frac{a_0 + a_2/3 + \ldots}{\omega_0} \,,
 \label{eq:par_lambdaB}
\end{eqnarray}
while the short-distance behavior in position space can be obtained from the Fourier transformed LCDA,
\begin{eqnarray}
	\tilde \phi_B^+(\tau,\mu_0) 
 &=& \int_0^\infty d\omega \, e^{-i \omega \tau } \, \phi_B^+(\omega,\mu_0)
 \cr 
   &=& \frac{1}{(1+i\omega_0\tau)^2}
	\, \sum_{k=0}^K a_k(\mu_0) \left( \frac{i\omega_0\tau-1}{i\omega_0\tau+1} \right)^k  
	\cr 
	&=& \sum_{k=0}^K (-1)^k a_k(\mu_0) \left[ 1  
   - 2 \, (k+1) \, i \omega_0\tau - (3+4k +2k^2) \, \omega_0^2\tau^2 + {\cal O}(i\omega_0\tau)^3 \right] \,.
   \cr &&
   \label{phitau_ak}
\end{eqnarray}
We also note that this parameterization allows for a straightforward implementation of the 1-loop renormalization-group evolution, using the eigenfunctions of the Lange-Neubert kernel \cite{Lange:2003ff} as derived in Ref.~\cite{Bell:2013tfa} and Ref.~\cite{Braun:2014owa}.

\subsection{Short-distance expansion}

The short-distance expansion (OPE) of the light-cone operator in the definition of the $B$-meson LCDA reads\footnote{Due to heavy-quark spin symmetry, the generic Dirac structure for the OPE of non-local currents reads $\bar q(\tau n) [\tau n,0] \Gamma h_v(0) = \sum \bar q(0) \, A(n,v,iD) \, \Gamma h_v(0)$.
Here the matrix $A(n,v,iD)$ will contain a string of an even number of $\gamma$ matrices in the massless case \cite{Lee:2005gza}. Linear terms in the spectator mass $m$, on the other hand, come with an \emph{odd} number of $\gamma$ matrices, as a consequence of the HQET Feynman rules, see also Appendix~\ref{sec:radiative-tail}.
}
(see also Ref.~\cite{Kawamura:2008vq} for the massless case)
\begin{eqnarray}
{\cal O}_+(\tau) &=&	\bar q(\tau n) \left[\tau n,0\right] \slashed n\gamma_5 \, h_v(0) = \sum_{n=3}^\infty \sum_{k=1}^{K_n} c_k^{(n)}(\tau) \, {\cal O}_k^{(n)}(0)
	\cr &=& c_{1}^{(3)}(\tau) \, \bar q \, \slashed n\gamma_5 \, h_v  
	\cr && {} + c_{1}^{(4)}(\tau) \, \bar q \, (i n \cdot \overleftarrow{D}) \, \slashed n\gamma_5 \, h_v
	  + c_{2}^{(4)}(\tau) \, \bar q \, (i v \cdot \overleftarrow{D}) \, \slashed n \gamma_5 \, h_v 
	   + c_{3}^{(4)}(\tau) \, m \, \bar q \, \slashed v \slashed n \gamma_5 \, h_v 
    \cr && {} + \ldots
    \label{ope} 
\end{eqnarray}
This translates into the following expression for the Fourier transformed LCDA,
\begin{eqnarray}
	\tilde\phi_+(\tau) =
    c_1^{(3)}(\tau)
    + \bar \Lambda \left(
        \frac43 \, c_1^{(4)}(\tau)
        + c_2^{(4)}(\tau)
    \right)
	- m \left(
         c_3^{(4)}(\tau)
         + \frac13 \, c_1^{(4)}(\tau)
     \right)
     + {\cal O}(\tau^2) \,,
 \label{phiexp}
\end{eqnarray}
where we have used the hadronic matrix elements of the local HQET
operators \cite{Grozin:1996pq,Bell:2008er},
\begin{eqnarray}
	\langle 0| {\cal O}_1^{(3)} | \bar{B}(v)\rangle 
	= i  m_B f_B^{\rm HQET}  \,,
\end{eqnarray}
and
\begin{eqnarray}
&&	\frac{\langle 0| {\cal O}_1^{(4)} |\bar{B}(v)\rangle}{\langle 0| {\cal O}_1^{(3)} | \bar{B}(v)\rangle} = \frac{4\bar\Lambda - m}{3} \,, 
	\quad 
	\frac{\langle 0| {\cal O}_2^{(4)} |\bar{B}(v)\rangle}{\langle 0| {\cal O}_1^{(3)} | \bar{B}(v)\rangle} = \bar\Lambda \,,
	\quad
	\frac{\langle 0| {\cal O}_3^{(4)} |\bar{B}(v)\rangle}{\langle 0| {\cal O}_1^{(3)} | \bar{B}(v)\rangle} = - m \,.
\end{eqnarray} 
Here $\bar\Lambda=m_B - m_b\big|_{\rm OS}$ refers to the HQET residual mass parameter in the on-shell scheme. In the following, we will renormalize the above operators in the $\overline{\mathrm{MS}}$ scheme, and the light quark mass $m=\overline{m}(\mu)$ is to be understood in the same scheme, accordingly.

The 1-loop contributions to the OPE coefficients for the massless case have already been calculated
in Ref.~\cite{Kawamura:2008vq}, with the result%
\footnote{In the same reference one can also find the 1-loop matching coefficients for the dimension-5 operators in the massless case.
A brief study of the effect of dimension-5 terms can be found in Appendix~\ref{sec:app-dim-5}.}
\begin{eqnarray}
	c_1^{(3)}(\tau) &=& 1 -\frac{\alpha_s C_F}{4\pi} 
	\left( 2 L^2+2 L + \frac{5\pi^2}{12}\right) + {\cal O}(\alpha_s^2) \,, \cr 
	c_1^{(4)}(\tau) &=& -i\tau \left[ 1 - \frac{\alpha_s C_F}{4\pi} \left( 2 L^2+L + \frac{5\pi^2}{12} \right)  + {\cal O}(\alpha_s^2) \right]  \,, \cr 
	c_{2}^{(4)}(\tau) &=& -i\tau \left[ - \frac{\alpha_s C_F}{4\pi} \left( 4L-3 \right) + {\cal O}(\alpha_s^2)\right]  \,,
\end{eqnarray} 
where $L= \ln \left( i\tau \mu e^{\gamma_E} \right)$ and $C_F=4/3$.
The operator ${\cal O}_3^{(4)}$ does not appear in the massless limit.
Its coefficient can be obtained in fixed-order perturbation theory by matching suitable
on-shell matrix elements of the left- and right-hand side of the OPE. 
A simple way is to consider partonic amplitudes between an incoming heavy quark with velocity $v^\mu$ and a light anti-quark with momentum $k^\mu=m v^\mu$ into the vacuum. 
This essentially corresponds to the setup in Ref.~\cite{Bell:2008er}, where the light-quark mass $m$ fulfills $m_b \gg m \gg \Lambda_{\rm QCD}$ and to first approximation the $B$-meson is described as a non-relativistic bound state. In this way one can infer the information on $c_3^{(4)}(\tau)$ by taking the Fourier transform of the 1-loop expression for the non-relativistic LCDAs and expanding to first order in the light quark mass $m$. Notice that in the non-relativistic limit the HQET parameter is identified with the light spectator mass, $$\bar\Lambda = m \left( 1 + {\cal O}(\alpha_s) \right) \,. $$
Therefore, the non-relativistic LCDAs only fix a linear combination of the dimension-4 contributions. Using the results for $c_1^{(3)}(\tau)$ and $c_{1,2}^{(4)}(\tau)$ from Ref.~\cite{Kawamura:2008vq}, we can determine the remaining coefficient $c_3^{(4)}(\tau)$ unambigously. As a by-product, we can also compute the ${\cal O}(\alpha_s)$ corrections to the relation $\bar\Lambda\simeq m$ in the non-relativistic set-up. In Appendix~\ref{sec:radiative-tail}, we will also present an independent derivation of the individual matching coefficients for an arbitrary Dirac structure, which includes the results for the short-distance expansion of ${\cal O}_+(\tau)$ as a special case.

\section{Matching coefficients for the massive case}

\label{sec:match}

We derive the matching relation by taking on-shell matrix elements of the left- and right-hand side of the OPE in Eq.~(\ref{ope}), 
and, for simplicity, projecting onto a particular spin configuration, such that for $k^\mu = m v^\mu$ the product of heavy- and light-quark
spinors can be replaced by
\begin{eqnarray*}
  u(v,s) \, \bar v(k,s') &\longrightarrow& - \frac{1+\slashed v}{2} \, \gamma_5 \,,
\end{eqnarray*}
up to an irrelevant normalization constant,
which amounts to setting
\begin{eqnarray*} 
&&  \bar v(k,s') \, \slashed n \gamma_5 \, u(v,s)  \longrightarrow  2 \,, 
  \qquad 
  \bar v(k,s') \, \slashed v \slashed n \gamma_5 \, u(v,s)  \longrightarrow  -2 \,.
\end{eqnarray*}
The calculation then corresponds to the non-relativistic setup in Ref.~\cite{Bell:2008er}, where the LCDAs of a $B$ meson with a massive light quark, $m_b \gg m \gg \Lambda_{\rm QCD}$, can be calculated in fixed-order perturbation theory.
With this one finds
\begin{eqnarray}
    && 1 - i m \tau + \frac{\alpha_s C_F}{4\pi} \, \tilde I^+(\tau)
    + \mathcal{O}(\tau^2, \alpha_s^2)
    \cr
    &=& c_1^{(3)}(\tau) \left(1 + \frac{\alpha_s C_F}{4\pi} \, \tilde I_1^{(3)} \right)
    + m \, c_1^{(4)}(\tau) \left(1 + \frac{\alpha_s C_F}{4\pi} \, \tilde I_1^{(4)}
        \right)
    \cr && {}
    + m \, c_2^{(4)}(\tau)
    - m \, c_3^{(4)}(\tau)
    + \mathcal{O}(\tau^2,\alpha_s^2) \,.
\end{eqnarray}
Here, we have already taken into account that the coefficients $c_{2,3}^{(4)}$ only start at order $\alpha_s$.
The function $\tilde I^+(\tau)$ on the left-hand side 
is the Fourier transform of the integral $I^+(\omega)$ which determines the ${\cal O}(\alpha_s)$ corrections to the non-relativistic LCDA 
as defined in Ref.~\cite{Bell:2008er}.
On the right-hand side, in the limit $\tau \to 0$,
the 1-loop corrections are encoded in the momentum-space integrals
\begin{equation}
    \tilde I_1^{(3)} = \int_0^\infty d{\omega}\, I^+(\omega)
    \quad \mathrm{and} \quad
    \tilde I_1^{(4)} = \frac{1}{m} \int_0^\infty d{\omega}\, \omega \, I^+(\omega) \,.
\end{equation}
The contribution of these diagrams is thus related to those of the non-local operator as given above, but we have to keep in mind that the $\overline{\mathrm{MS}}$ subtraction has to be performed \emph{after} the $\omega$ integration, in order to account for the renormalization of \emph{local} operators.
To the considered order, the matching relation can thus be rewritten as 
\begin{eqnarray}
   &&\frac{\alpha_s C_F}{4\pi} \left( \tilde I^+(\tau) - \tilde I_1^{(3)} + i m \tau \, \tilde I_1^{(4)} \right)
    \cr 
    &=&  \left( c_1^{(3)}(\tau) - 1 \right) + m \left( i\tau + c_1^{(4)}(\tau)+
    c_2^{(4)}(\tau) - c_3^{(4)}(\tau) \right)
        + \mathcal{O}(\tau^2, \alpha_s^2) \,.
        \label{match}
\end{eqnarray}

\subsection{Analysis of the individual diagrams in Feynman gauge}

\begin{figure}[t!]
\begin{center} 
\includegraphics[width=0.8\textwidth]{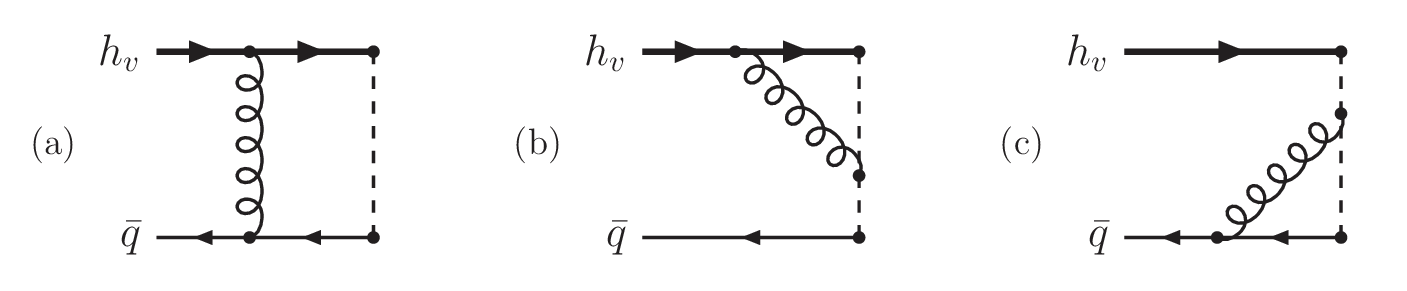}
\end{center}
 \caption{\label{fig:1loop} The three 1-loop Feynman diagrams contributing to the matching calculation. (The dashed line indicates the Wilson line.)}
\end{figure}
It is instructive to consider the individual contributing Feynman diagrams (in Feynman gauge),  as illustrated in Fig.~\ref{fig:1loop}.

\subsubsection*{(a) Vertex correction}

Let us first consider the vertex correction, i.e.\ the gluon exchange between the light and heavy quark. Before $\overline{\rm MS}$ subtraction, the result for the left-hand side of the OPE in momentum space reads \cite{Bell:2008er}
\begin{eqnarray}
	I_a^+(\omega) &=& 2\omega \, \Gamma(1+\epsilon) 
	\left( \frac{\mu^2 e^{\gamma_E}}{(m-\omega)^2} \right)^\epsilon 
	\left\{ \frac{2}{(m-\omega)^2} - \frac{\theta(m-\omega)}{m \, (m-\omega)} - \frac{\theta(\omega-m)}{\omega\,(\omega-m)}\right\}
\end{eqnarray}
From this we calculate the Fourier transform as 
\begin{eqnarray} 
  &&\tilde I_a^+(\tau) 
  =  \int_0^\infty d\omega \, e^{ - i \omega \tau} \, I_a^+(\omega) 
   \cr 
   &=& e^{-im\tau} \left( \frac{2}{\epsilon} + 4 L - 
   6 \, {\rm Ei}(im\tau) + 4 i m\tau \, {\rm Ei}(im\tau) \right)
       -4 
       + \frac{2-2 \, e^{-im\tau}}{im\tau} 
       + {\cal O}\left(\epsilon \right)
       \cr 
   &=&  \frac{2}{\epsilon} - 2 L + 3 \, \ln \frac{\mu^2}{m^2} - 2 
   - i \tau m \left( \frac{2}{\epsilon} -6 L + 5 \, \ln \frac{\mu^2}{m^2} + 7 \right) 
   + {\cal O}(m^2,\epsilon)
\end{eqnarray}
where $L$ is defined as above and $\mathrm{Ei}(z)$ is the exponential integral function.
Similarly, the vertex correction to the local operator ${\cal O}_1^{(3)}$ is obtained as \cite{Bell:2008er}
\begin{eqnarray}
  &&  \tilde I_{1,a}^{(3)} = \int_0^\infty d\omega \, I_a^+(\omega) 
  = \frac{3}{\epsilon} + 3 \, \ln \frac{\mu^2}{m^2} -2 + {\cal O}(\epsilon) \,.
\end{eqnarray}
The vertex correction to the local operator ${\cal O}_1^{(4)}$ 
translates into the $\omega$ moment of the integral $I_a^+(\omega)$,
\begin{eqnarray}
 &&    \tilde I_{1,a}^{(4)} = \frac{1}{m} \,
 \int_0^\infty d\omega \, \omega  \, I_a^+(\omega) = 
 \frac{5}{\epsilon} + 5 \, \ln \frac{\mu^2}{m^2} + 3 + {\cal O}(\epsilon) \,.
\end{eqnarray}
With this the contribution of the vertex correction to the matching relation is obtained as
\begin{eqnarray}
 && \tilde I_a^+(\tau) - \tilde I_{1,a}^{(3)} + im\tau \, \tilde I_{1,a}^{(4)} 
 \cr 
 &=& \int_0^\infty d\omega \left( e^{-i\omega \tau} - 1 +i \omega \tau   \right) I_a^+(\omega)
  \cr
  &=& - \frac{1}{\epsilon} - 2 L  - i\tau m \left( -\frac{3}{\epsilon} -6 L + 4 \right) + {\cal O}(m^2,\epsilon)
  \label{eq:Iatilde}
\end{eqnarray} 
Notice that the dependence on the IR logarithms $\ln \frac{\mu^2}{m^2}$ 
has dropped out in the difference of the three terms, such that the matching coefficients only depend on the UV logarithms $L$, as they should.
Furthermore, note that the above relation holds for the \emph{bare} Feynman integral, and the final (and finite) result stems from the fact that the 
short-distance expansion of the exponential $e^{-i\omega\tau}$ and the expansion in $\epsilon$ of the dimensionally regularized integrals do not commute.
On the other hand, the subtraction terms generated by the expansion of the Fourier factor,
$$
 e^{-i \omega \tau} \to e^{-i \omega \tau} - \sum_n \frac{(-i\omega \tau)^n}{n!} 
$$
allow us to expand the remaining integrand $I_a^+(\omega)$ in momentum space with respect to the small mass parameter $m$ \emph{prior} to the integration,
\begin{eqnarray} 
  I_a^+(\omega) &\simeq & \Gamma(1+\epsilon) \left( \frac{\mu^2 e^{\gamma_E}}{\omega^2} \right)^\epsilon \left( \frac{2}{\omega} + (6+4\epsilon) \, 
   \frac{m}{\omega^2} + {\cal O}(m^2/\omega^3)\right) \,,
\end{eqnarray}
which reflects the contribution to the radiative tail of the LCDA in momentum space.
We have used this strategy to compute the matching contribution of the vertex diagram for a generic strange-quark momentum in Appendix~\ref{sec:radiative-tail}.

\subsubsection*{(b) Wilson line with heavy quark}

Next, we consider the coupling of the Wilson-line gluon to the heavy quark which yields
\cite{Bell:2008er} 
\begin{eqnarray}
	I_b^+(\omega) &=& 2 \, \Gamma(\epsilon) \, \int_0^\infty dk 
	\left( \frac{\mu^2 e^{\gamma_E}}{k^2} \right)^\epsilon 
	\frac{\delta(\omega-m-k)-\delta(\omega-m)}{k} \,.
\end{eqnarray}
The Fourier transform reads
\begin{eqnarray}
	\tilde I_b^+(\tau) &=& \int_0^\infty \, d\omega \, e^{-i\omega\tau} \, I_b^
	+(\omega) 
	\cr 
   &=& e^{-i\tau m}
	\left( -\frac{1}{\epsilon^2}
	- \frac{2L}{\epsilon}
	- 2 L^2 - \frac{5\pi^2}{12} \right) 
	+ \mathcal{O}(\epsilon) \,.
    \label{eq:Ibtilde}
\end{eqnarray}
In this case, there are no local subtraction integrals,
\begin{eqnarray} 
\int_0^\infty \, d\omega \left(1 - i \omega\tau  \right) I_b^
	+(\omega) 
  &=& 2 \, \Gamma(\epsilon) \, \int_0^\infty dk 
	\left( \frac{\mu^2 e^{\gamma_E}}{k^2} \right)^\epsilon
 \left( -i \tau  \right) = 0 \,,
\end{eqnarray}
because any term from the Taylor expansion of the exponential results in scaleless integrals in dimensional regularization.
As a consequence, the short-distance expansion for $\tilde I_b^+(\tau)$ in Eq.~(\ref{eq:Ibtilde}) factorizes into the trivial expansion of the Fourier factor and a \emph{universal} 
1-loop factor which includes the double-logarithmic dependence on $\tau$ and affects all tree-level matching coefficients in the same way.

\subsubsection*{(c) Wilson line with light quark}

Finally, the diagram with the Wilson-line gluon coupling to the light quark reads
\cite{Bell:2008er}
\begin{eqnarray}
	I_c^+(\omega)&=& 2 \Gamma(\epsilon) \, \int_0^m dk \, \frac{m-k}{m} \left(\frac{\mu^2 e^{\gamma_E}}{k^2}\right)^\epsilon \frac{\delta(k-m+\omega)-\delta(\omega-m)}{k} \,,
\end{eqnarray}
where the Fourier transform is given by
\begin{eqnarray}
		\tilde I_c^+(\tau)  &=& 
  \int_0^\infty \, d\omega \,  e^{-i\omega\tau} \,  I_c^+(\omega)
  \cr 
	&=& i\tau m \left( \frac{1}{\epsilon} + \ln \frac{\mu^2}{m^2} + 3  \right) + {\cal O}\left( (i\tau m)^2,\epsilon\right) \,.
\end{eqnarray}
The analogous contribution from the local operator yields $\tilde I_{1,c}^{(3)} = 0$ and
\begin{eqnarray}
	\tilde I_{1,c}^{(4)} &=& - \frac{2 \Gamma(\epsilon)}{m} \, \int_0^m dk \, \frac{m-k}{m} \left(\frac{\mu^2 e^{\gamma_E}}{k^2}\right)^\epsilon \cr
    &=& \left(\frac{\mu^2 e^{\gamma_E}}{m^2}\right)^\epsilon
        \frac{\Gamma(\epsilon - 1)}{2 \epsilon - 1}
    = - \bigg(\frac{1}{\epsilon} + 3 + \ln \frac{\mu^2}{m^2}\bigg) + {\cal O}(\epsilon)  \,,
\end{eqnarray}
As the integral $I_c^+(\omega)$ only involves the low-momentum region, $\omega <m$, the 
short-distance expansion of the Fourier exponential and dimensional regularization commute, 
and therefore the net contribution to the matching from diagram $(c)$ is zero,
\begin{eqnarray}
    \tilde I_c^+(\tau) - \tilde I_{1,c}^{(3)} + im\tau \, \tilde I_{1,c}^{(4)} &=& 
  \int_0^\infty \, d\omega \left( e^{-i\omega\tau} - 1 + i \omega \tau  \right)  I_c^+(\omega) =0
  \,.
\end{eqnarray}

\subsection{1-loop result for the matching coefficient \texorpdfstring{$c_3^{(4)}(\tau)$}{c34}}

Inserting the above results for the 1-loop integrals and the known Wilson coefficients from the massless case into the matching relation (\ref{match}), we obtain the following result for the remaining Wilson coefficient $c_3^{(4)}(\tau)$ after $\overline{\mathrm{MS}}$ renormalization:
\begin{eqnarray}
    c_3^{(4)}(\tau) &=& -i \tau \left[ \frac{\alpha_s C_F}{4\pi} \left( L -1 \right) + {\cal O}(\alpha_s^2) \right] \,,
    \label{eq:c34plus}
\end{eqnarray}
which represents one of the new theoretical results of our analysis.

\subsection{1-loop result for \texorpdfstring{$\bar\Lambda$}{LambdaBar} in the non-relativistic limit}

As a cross-check and by-product, we can insert the result for the Wilson coefficients into the OPE result for the LCDA in Eq.~(\ref{phiexp}), and obtain
\begin{eqnarray}
    \tilde\phi_+(\tau) &=& 
    \left[1- i \tau \, \frac{4 \bar\Lambda  - m }{3} \right]\left[  1 - \frac{\alpha_s C_F}{4\pi} \left( 2 L^2+2 L + \frac{5\pi^2}{12} \right) \right]
    \cr 
    && {} + i \tau \bar\Lambda \, \frac{\alpha_s C_F}{4\pi} \left( \frac83 \, L -3 \right) 
      + i \tau m \, \frac{\alpha_s C_F}{4\pi} \left( \frac43 \, L - 1 \right)  + {\cal O}(\alpha_s^2,\tau^2)
    \label{phitauexp}
\end{eqnarray}
On the other hand, the direct computation of the 1-loop corrections to the non-relativistic LCDA in position space 
results in (see also Ref.~\cite{Bell:2008er} for the momentum-space computation)
\begin{eqnarray}
    \tilde\phi_+(\tau) \Big|_{\rm NR} &=& 1 - i \tau m+ \frac{\alpha_s C_F}{4\pi} \left( 
    \tilde I_a^+(\tau) + \tilde I_b^+(\tau) + \tilde I_c^+(\tau) - \tilde I_{1,a}^{(3)} \right)_{\overline{\mathrm{MS}}}  + \ldots 
\end{eqnarray}
Comparing the two expressions,
we can read off the relation between the HQET parameter $\bar\Lambda$ and the light quark mass in the non-relativistic limit,
\begin{eqnarray}
    \bar\Lambda \Big|_{\rm NR}&=& m \left[ 1 + 3 \, \frac{\alpha_s C_F}{4\pi} \, \ln \frac{\mu^2}{m^2} + {\cal O}(\alpha_s^2)\right] \,.
\end{eqnarray}
With $m = \overline{m}(\mu)$ in the $\overline{\mathrm{MS}}$ scheme, and $\bar \Lambda=\bar\Lambda_{\rm pol}$ in the pole-mass scheme, one indeed finds
$d\bar\Lambda/d\ln\mu=0$ to the considered order in $\alpha_s$.

\section{Constraints on the generic LCDA parameterization}
\label{sec:num}

In this section we will work out the theoretical constraints on the $B$-meson LCDAs that follow from the perturbative results for the radiative tail, together with the generic parameterization in Eq.~(\ref{eq:generic}),
following the procedure outlined in Ref.~\cite{Feldmann:2022uok}.
The main purpose of this analysis is to determine whether the theoretical information resulting from the radiative tail is compatible with complementary studies of the inverse moments of the $B$-meson LCDAs from QCD sum rules. Indeed, we find very good agreement, both on a qualitative and quantitative level (within the uncertainties).
We stress that the results from the perturbative tail \emph{alone} cannot provide precise predictions for the LCDAs.
Rather, the conclusion to be drawn is that the constraints could and should be used -- together with other independent theoretical information
from sum rules or lattice -- in future phenomenological analyses of $B_s$ decays in the framework of QCD factorization or QCD light-cone sum rules.

\subsection{Determination of the expansion parameters}

Following Ref.~\cite{Feldmann:2022uok}, the short-distance expansion for the LCDA $\tilde\phi_+(\tau)$ in Eq.~(\ref{phitauexp}) translates into constraints onto the expansion coefficients $a_k$ in Eq.~(\ref{phitau_ak}).
To this end, we take an imaginary-valued reference point $\tau=\tau_0$ such that 
$$ 
x_0 \equiv i \tau_0 \mu_0 \, e^{\gamma_E} = {\cal O}(1) \,,
$$
for a given reference scale, which we will fix as $\mu_0=1$~GeV. This ensures that the logarithms $L=\log x_0$ 
in the matching coefficients of the OPE are not large. In the numerical analysis below, we fix $x_0\equiv 1$ for simplicity.
In order to compare with our parameterization, we have to expand $\tilde \phi_B^+(\tau)$ in powers of $\tau_0$,
and therefore for this expansion to converge we further have to require that
the auxiliary reference momentum $\omega_0$ 
in the parameterization of the $B_s$ or $B_q$ LCDA satisfies
$$
n_0 \equiv i \tau_0 \omega_0 \ll 1 \,.
$$
Finally, as becomes apparent below, 
we have to require $\omega_0 \gtrsim \bar\Lambda_a, m_a$ to avoid large enhancement factors in the 
resulting expressions for the expansion parameters $a_k$. In the numerical analysis, we use $n_0\equiv 1/3$
which satisfies these requirements.

In the following, we consider the parameterization of the LCDA truncated at $K=2$. The OPE results for the LCDA and its first derivative at the point $\tau_0$ provide two independent conditions that determine the parameters $a_0$ and $a_1$, while $a_2$ remains unconstrained.
In this way, we find
\begin{eqnarray}
    a_0 &=& 
    2 + a_2 - \frac{4\bar\Lambda- m}{6\omega_0} + \frac{\alpha_s C_F}{4\pi} \left( - \frac{1}{x_0} \, \frac{\mu_0 \mathrm{e}^{\gamma_E}}{\omega_0} \, (1+ 2 \ln x_0 ) + \ldots \right) \,, 
    \\
    a_1 &=& 1 + 2 a_2 - \frac{4\bar\Lambda- m}{6\omega_0} + \frac{\alpha_s C_F}{4\pi} \left(
     - \frac{1}{x_0} \, \frac{\mu_0 \mathrm{e}^{\gamma_E}}{\omega_0} \, (1+ 2 \ln x_0 ) + \ldots\right) \,, 
\end{eqnarray}
where only the $\alpha_s$ corrections that are enhanced by $\mu_0 / \omega_0$ are shown for the moment. 
As those are independent of the light quark mass, it is convenient to absorb them by the same redefinition as in Ref.~\cite{Feldmann:2022uok},
\begin{eqnarray}
    \bar \Lambda & \equiv  &
    \bar \Lambda_a(\mu_0, x_0)
    \left[ 1 + \frac{\alpha_s C_F}{4\pi} \left( 
    10\, \ln x_0 + \frac{15}{4} \right) \right]
    - \frac{\alpha_s C_F}{4\pi}  \, \frac{3\mu_0 \mathrm{e}^{\gamma_E}}{2x_0} \left( 1+ 2 \ln x_0 \right) 
    \label{eq:Lama1} \,.
\end{eqnarray}
With this definition, the 1-loop result for the expansion parameters with $K=2$ reads
\begin{eqnarray}
    a_0  &=& Z(x_0) \left( 2 - 
    \frac{4 \bar\Lambda_a(\mu_0, x_0) - m_a(x_0)}{6\omega_0}
    - 2 \, r(x_0)
    \right) + a_2 \,, 
    \cr
    a_1  &=& Z(x_0) \left( 1 - 
    \frac{4 \bar\Lambda_a(\mu_0, x_0) - m_a(x_0)}{6\omega_0}
    - r(x_0) 
    \right) +2 \, a_2
    \,,
    \label{eq:exp_coeff_res}
\end{eqnarray}
which holds to order ${\cal O}(\alpha_s)$. Here we  introduce 
\begin{eqnarray}
    r(x_0) &\equiv& \frac{\alpha_s C_F}{6\pi} \, \frac{8\bar\Lambda_a \, x_0 (1+\ln x_0) + m_a \, x_0 (1- 2\ln x_0)}{\mu_0 \mathrm{e}^{\gamma_E}} = {\cal O}(\alpha_s n_0) \,,
\end{eqnarray}
and 
\begin{eqnarray}
    Z(x_0) &\equiv& 1+\frac{\alpha_s C_F}{4\pi} \left( -2\ln^2 x_0+2\ln x_0 + 2 - \frac{5\pi^2}{12} \right) \,,
\end{eqnarray}
and 
\begin{eqnarray}
    m_a(x_0) &\equiv & m \left( 1 - \frac{\alpha_s C_F}{4\pi} \left( 3+4 \ln x_0 \right) \right) \,,
    \label{eq:madef}
\end{eqnarray}
as a short-hand notation.
Our definitions of $\bar\Lambda_a$, $m_a$ and $Z$  have been chosen such that the parameterization for the position-space LCDA with finite truncation $K$ satisfies
\begin{eqnarray}
 \tilde\phi_+(0)\big|_K &=& \sum_{k=0}^K (-1)^k \, a_k = Z(x_0) - r(x_0)
 + {\cal O}(\alpha_s^2,\, n_0^2) \,,
 \cr 
  \tilde\phi_+'(0)\big|_K &=& - 2 i \omega_0 \, \sum_{k=0}^K (-1)^k \, (1+k) \, a_k
  = - Z(x_0) \, \frac{4 i \bar \Lambda_a- i m_a}{3} + {\cal O}(\alpha_s^2,\, \bar \Lambda_a n_0) \,,
  \label{eq:Lama_scheme}
\end{eqnarray}
which generalizes the corresponding relations in Ref.~\cite{Feldmann:2022uok} to the case of a massive light spectator quark.

\subsection{Numerical results}

As stressed in the previous subsection, the aim of the numerical analysis is \emph{not} to provide precise predictions for 
the $B_q$ and $B_s$-meson LCDAs, but rather to figure out to what extent the theoretical constraints from 
the radiative tail can be used in future phenomenological analyses. For that reason, our focus will be on the inverse moments $\lambda_{B_q}$ and $\lambda_{B_s}$, which play a dominant role in applications of QCD factorization to exclusive $B$ decays. The plots that we show below are aimed to illustrate 
our findings on a semi-quantitative level. For that reason, most of the input parameters are simply set to their central values.
Nevertheless, we quote the expected uncertainties on the input parameters, where possible.

As already stated, we take $x_0\equiv 1$ and $n_0=1/3$ for the dimensionless combinations of $\tau_0$, $\mu_0$ and $\omega_0$ as defined in the previous section.
The renormalization scale is fixed to
$\mu_0=1~{\rm GeV}$, which results in $\omega_0 \simeq 594$~MeV.
The corresponding value of the strong coupling is taken as $\alpha_s(\mu_0)=0.5$. With this the value of the HQET parameter for the $B_q$ meson in the scheme defined above is \cite{Feldmann:2022uok}
$$
\bar\Lambda_a^{(q)}(\mu_0, x_0=1) \simeq 367~{\rm MeV} \,. 
$$
The corresponding value for $B_s$ mesons is obtained by taking the hadronic mass
differences from $M_{B_s} = 5.367~\mathrm{GeV}$, $M_{B_q} = 5.279~\mathrm{GeV}$ \cite{ParticleDataGroup:2022pth},
together with Eq.~(\ref{eq:Lama1}), leading to 
$$
\bar\Lambda_a^{(s)}(\mu_0,1) \simeq \bar\Lambda_a^{(q)}(\mu_0, 1) 
    + \left(1-\frac{\alpha_s C_F}{4\pi} \, \frac{15}{4} \right) (M_{B_s}-M_{B_q}) \simeq 437~\mathrm{MeV} \,.
$$
For the strange-quark mass in the $\overline{\rm MS}$ scheme we adopt
$m_s(\mu_0)= 126^{+15}_{-7}~\mathrm{MeV}$ \cite{ParticleDataGroup:2022pth},
which via Eq.~(\ref{eq:madef}) translates to 
$$
\quad m_a^{(s)}(\mu_0) \simeq (106 \pm 10)~\mathrm{MeV} \,.
$$

\subsubsection*{Inverse moment of the LCDA for $B_q$ meson}

\begin{figure}[t!p]
\begin{center}
    \includegraphics[scale=0.8]{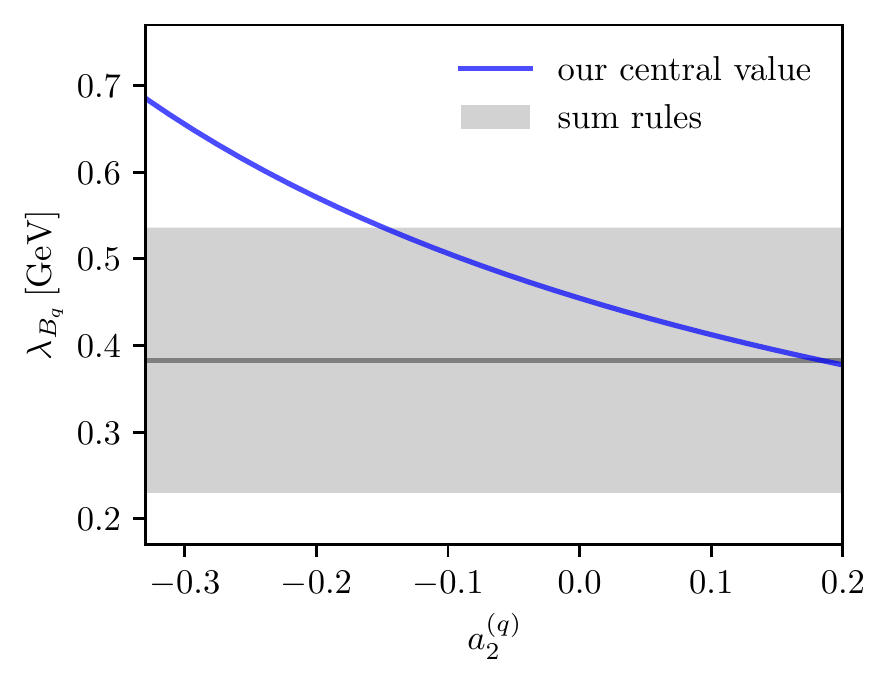} \quad
    \includegraphics[scale=0.8]{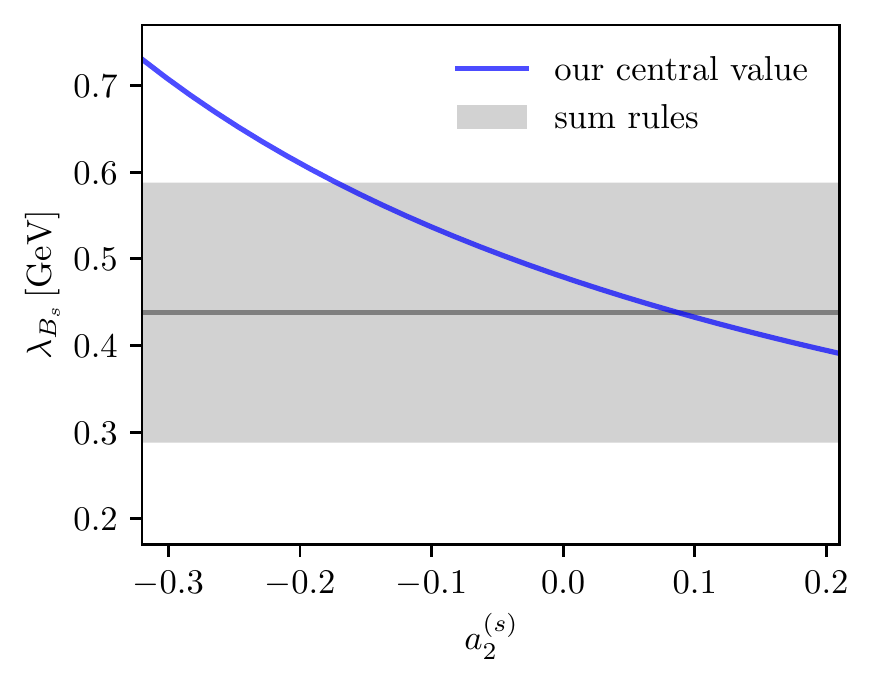}
\end{center}
\caption{\label{fig:numerical-results-inverse-moments}
     Estimates for the inverse moment $\lambda_{B}$ of the $B$-meson LCDA (Left: for $B_q$. Right: for $B_s$). 
     Blue solid line: central value following from the 
        constraint of the radiative tail on the truncated parameterization, as a function of the expansion coefficient $a_2$. 
        Gray band: estimate from the latest sum-rule analysis \cite{Khodjamirian:2020hob}.
}
\end{figure}

Inserting the above values into the matching relations for the expansion coefficients Eq.~(\ref{eq:exp_coeff_res}), we find for the $B_q$ meson
\begin{eqnarray}
    a^{(q)}_0 &\simeq& 1.78 - 0.47 \, \frac{\bar \Lambda_a^{(q)}}{367~{\rm MeV}} + a_2^{(q)} \simeq 1.31 + a^{(q)}_2 \,, \cr 
    a^{(q)}_1 &\simeq& 0.89 - 0.42 \, \frac{\bar \Lambda_a^{(q)}}{367~{\rm MeV}} + 2 a_2^{(q)} \simeq 0.47 + 2 a^{(q)}_2 \,.
\end{eqnarray}
To establish upper and lower limits for $a^{(q)}_2$, we adopt the ad-hoc criterion suggested in Ref.~\cite{Feldmann:2022uok},
\begin{eqnarray}
    \frac{ \big| a^{(q)}_1 \big|^2 }{ \big| a^{(q)}_0 \big|^2 + \big| a^{(q)}_1 \big|^2 } < 0.25 \,, \qquad 
    \frac{ \big| a^{(q)}_2 \big|^2 }{ \big| a^{(q)}_0 \big|^2 + \big| a^{(q)}_1 \big|^2 + \big| a^{(q)}_2 \big|^2 } < 0.1 \,,
\label{eq:crit}
\end{eqnarray}
which should ensure a reasonable convergence of the integral bound in Eq.~(\ref{eq:bound}).
Numerically, this translates to
\begin{eqnarray}
    -0.33 < a^{(q)}_2 < 0.20 \,.
    \label{eq:critq}
\end{eqnarray}
The estimate for the inverse moment from Eq.~(\ref{eq:par_lambdaB}),
\begin{eqnarray}
 \lambda_{B_q} &\simeq & \frac{445~{\rm MeV}}{1.33 - 0.35 \, \frac{\bar\Lambda_a^{(q)}}{367~{\rm MeV}} + a_2^{(q)}} 
 \simeq \frac{445~{\rm MeV}}{0.98 + a_2^{(q)}} \,,
\end{eqnarray}
is depicted in Fig.~\ref{fig:numerical-results-inverse-moments} in the given interval for the expansion parameter $a_2^{(q)}$. 
We compare this estimate with the latest sum-rule result $\lambda_{B_q} = 383\pm153~\mathrm{MeV}$ \cite{Khodjamirian:2020hob} 
and observe that, for a large region of the considered $a_2^{(q)}$ interval, both estimates are in good agreement,
with a slight preference for positive values of $a_2^{(q)}$. 
The value of $a_2^{(q)}$ can also be roughly estimated by adhering to $K=2$ and including the constraints from 
the dimension-5 operators which is studied in Appendix~\ref{sec:app-dim-5} and yields compatible values 
with the above analysis within the uncertainties.
We also notice that the value for the inverse moment $\lambda_{B_q}$ 
is positively correlated with the value of the HQET parameter $\bar\Lambda_a^{(q)}$, 
but the two hadronic quantities are not simply proportional to each other.

\subsubsection*{Inverse moment of the LCDA for $B_s$ meson}

Similarly, the coefficients for the $B_s$ meson are calculated as
\begin{eqnarray}
    a^{(s)}_0 &\simeq& 1.78
        - 0.56 \, \frac{\bar\Lambda_a^{(s)}}{437~\mathrm{MeV}}
        + 0.023 \, \frac{m_a^{(s)}}{105~\mathrm{MeV}}
        + a^{(s)}_2
        \simeq 1.24 + a^{(s)}_2 \,,\cr
    a^{(s)}_1 &\simeq& 0.89
        - 0.50 \, \frac{\bar\Lambda_a^{(s)}}{437~\mathrm{MeV}}
        + 0.025 \, \frac{m_a^{(s)}}{106~\mathrm{MeV}}
        + 2 a^{(s)}_2
        \simeq 0.42 + 2 a^{(s)}_2
        \,.
\end{eqnarray}
Compared to the $B_q$ meson, the resulting flavor-symmetry breaking effect
for the coefficients $a_{0,1}^{(s)}$ is of the expected size (10-15\%).
It is thus reasonable to expect that this remains true for the 
(yet) undetermined coefficients $a_2^{(q,s)}$ as well.
Moreover, we can use the analogous convergence criterion in Eq.~(\ref{eq:crit}) as for the $B_q$ meson 
to constrain the interval to consider for the parameter $a_2^{(s)}$,
leading to 
\begin{eqnarray}
    -0.32 < a^{(s)}_2 < 0.21 \,.
    \label{eq:crits}
\end{eqnarray}
Considering now the inverse moment of the $B_s$-meson LCDA, we find
\begin{eqnarray}
    \lambda_{B_s} &\simeq&
        \frac{455~\mathrm{MeV}}{
            1.33 
            - 0.42 \frac{\bar\Lambda_a^{(s)}}{437~\mathrm{MeV}}
            + 0.017 \, \frac{m_a^{(s)}}{106~\mathrm{MeV}}
            + a^{(s)}_2
            }
        \simeq \frac{455~\mathrm{MeV}}{0.93 + a^{(s)}_2} \,.
\end{eqnarray}
The aforementioned correlation between $\lambda_B$ and $\bar \Lambda_a$ remains, while 
the explicit effect of the strange-quark mass turns out to be marginal.
On the right-hand side of Fig.~\ref{fig:numerical-results-inverse-moments} we show our result 
as a function of $a_2^{(s)}$ in comparison with the value $\lambda_{B_s} = 438 \pm 150~\mathrm{MeV}$,
which was determined from QCD sum rules \cite{Khodjamirian:2020hob}.
We find again that our approach to implement the constraints from the radiative tail is well compatible with the 
sum-rule estimates.

\subsubsection*{The ratio $\lambda_{B_s}/\lambda_{B_q}$}

\begin{figure}[t!p]
\begin{center}
    \includegraphics[scale=0.8]{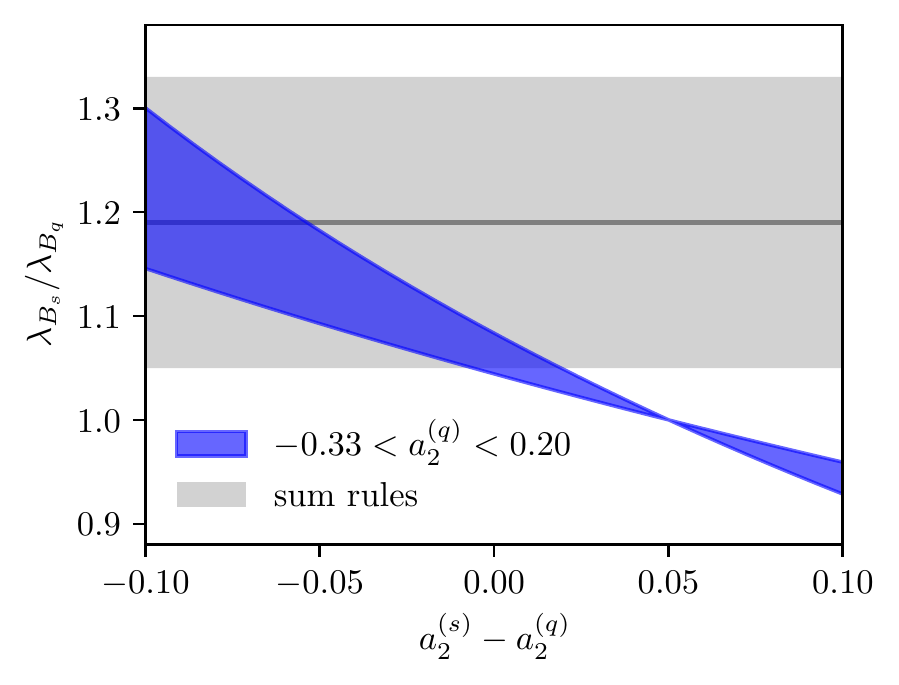}\quad
    \includegraphics[scale=0.8]{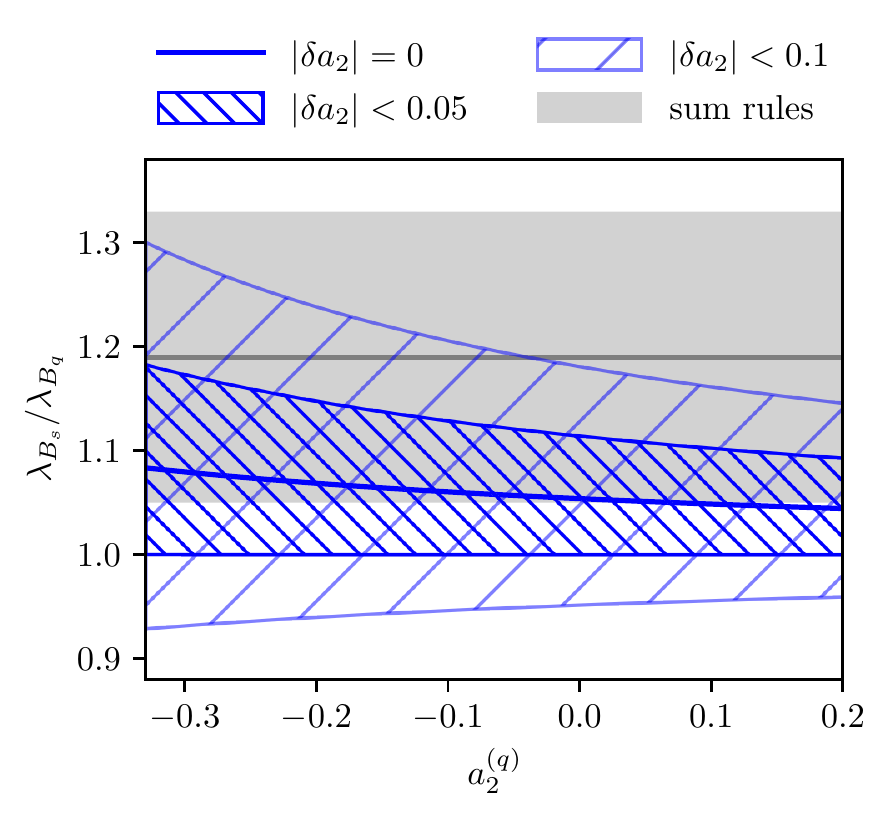}
\end{center}
\caption{\label{fig:inverse-moment-ratio}
    Estimates for the ratio of inverse moments $\lambda_{B_s} / \lambda_{B_q}$ of the $B_{s,q}$-meson LCDAs
    (Left: as a function $a^{(s)}_2 - a^{(q)}_2 \equiv \delta a_2$. Right: as a function of $a^{(q)}_2$). 
    Blue (hatched) bands: value ranges following from the 
    constraint of the radiative tail on the parameterization, under the given additional constraint.
    Gray band: estimate from the latest sum-rule analysis \cite{Khodjamirian:2020hob}.
}
\end{figure}

Considering the ratio of inverse moments, we find
\begin{eqnarray}
    \frac{\lambda_{B_s}}{\lambda_{B_q}} &\simeq&
        \frac{
            1.33 - 0.35 \, \frac{\bar\Lambda^{(q)}_a}{367~\mathrm{MeV}} + a^{(q)}_2
        }{
            1.33 - 0.42 \, \frac{\bar\Lambda^{(s)}_a}{437~\mathrm{MeV}}
            + 0.017 \, \frac{m^{(s)}_a}{106~\mathrm{MeV}}
            + a^{(s)}_2
        }
        \simeq \frac{0.98 + a^{(q)}_2}{0.93 + a^{(s)}_2} \,.
        \label{eq:ratio}
\end{eqnarray}
As already discussed, the main flavor-symmetry breaking effect stems from
the difference between the HQET parameters for $B_q$ and $B_s$ mesons, while
the explicit effect of the strange-quark mass is small. 
This leaves the dependence on the undetermined coefficients $a_2^{(q,s)}$ which
are varied on a compact parameter space, 
constrained by the convergence of the integral bound and flavor symmetry.
Given that the maximal values for $a_2^{(q,s)}$ allowed by our convergence criterion are about 0.3, 
and we do not expect flavor-symmetry corrections to be
larger than 30\%, we consider $|\delta a_2|< 0.1$ to be a conservative bound.
In Fig.~\ref{fig:inverse-moment-ratio} we illustrate our numerical result for Eq.~(\ref{eq:ratio}) in two 
different ways:
On the left-hand side we plot the ratio as a function of the difference $\delta a_2 = a_2^{(s)}-a_2^{(q)}$, with 
the value of $a_2^{(q)}$ varied within the interval constrained by Eq.~(\ref{eq:crit}). On the right-hand side, we plot 
the ratio as a function of $a_2^{(q)}$, assuming different ranges for $|\delta a_2|$.
Again, in both cases we observe good agreement with the sum-rule estimate $1.19\pm 0.14$ 
for that ratio from Ref.~\cite{Khodjamirian:2020hob} (which has smaller
uncertainty than the estimates for the individual inverse moments). While $\delta a_2=0$ is allowed, the comparison shows
a slight preference for $a_2^{(s)}-a_2^{(q)} <0$, which would repeat the trend seen in $a_{0,1}^{(s)}-a_{0,1}^{(q)}$,
and 
is also in line with the constraints from the dimension-5 operators provided in Appendix~\ref{sec:app-dim-5}.

In conclusion, our numerical study reveals that the perturbative constraints from the radiative tail of the LCDAs, 
together with our generic parameterization, yield inverse moments that are consistent with independent studies, 
even at a low truncation order. 
Our formalism shows a clear correlation between the inverse moments of the LCDAs and the HQET parameter $\bar\Lambda$, 
where in most parts of the allowed parameter space we have $\lambda_{B_s}>\lambda_{B_q}$. We also find that the explicit effect of the
strange-quark mass in the short-distance expansion gives only a minor effect on our result. For that reason, we have also
extrapolated our analysis to the case of the $B_c$ meson in HQET (assuming $m_c \sim \mu_0 \ll m_b$), which
is briefly summarized in Appendix~\ref{sec:Bc}.

\clearpage

\section{Summary}

To summarize, in this work we have studied the 2-particle light-cone distribution amplitudes (LCDAs) of the $B_s$ meson, focusing on the effect
of the strange-quark mass on the radiative tail, i.e.\ the behavior of the LCDAs at large light-cone momentum fraction 
or small light-cone separations. To this end, we have reconsidered the short-distance expansion of 2-particle
light-ray operators in HQET, and calculated the 1-loop matching coefficients, 
where our new results include the dimension-4 operator proportional to the strange-quark mass, see Eq.~(\ref{eq:c34plus}), and 
the case of the subleading 2-particle $B$-meson LCDA, see Eq.~(\ref{eq:d34minus}).
We have shown in some detail how the matching procedure can be conveniently performed, starting from the asymptotic behavior of 
momentum-space Feynman integrals for on-shell matrix elements, 
and performing the necessary local subtractions on the level of the Fourier transform to position space, 
prior to the $\overline{\rm MS}$ subtractions, see the discussion after Eq.~(\ref{eq:Iatilde}). 
In this way, our approach is very similar to the calculation of the LCDAs for non-relativistic bound states in HQET \cite{Bell:2008er}.

On the basis of this result, we have studied the numerical effect of the short-distance constraints on the level of a generic
parameterization of the LCDA, proposed by two of us in Ref.~\cite{Feldmann:2022uok}. The value and first derivative of the LCDA in
position space at a suitably chosen small but non-zero light-cone separation are used to constrain two expansion coefficients.
The size of the remaining coefficients is limited by an integral bound. We have used this approach to obtain numerical estimates
for inverse moment of the leading $B_q$ and $B_s$ LCDAs, which turn out to be in very reasonable agreement with independent
results from QCD sum rules \cite{Khodjamirian:2020hob}, see Figs.~\ref{fig:numerical-results-inverse-moments} and \ref{fig:inverse-moment-ratio}.
In particular, we find that the dominant flavor-symmetry breaking effect for the radiative tail is induced by the difference
between the HQET mass parameters, $\bar\Lambda_{B_s} > \bar \Lambda_{B_q}$, while the explicit effect of the strange-quark mass 
only has a minor numerical effect. This suggests that our result can also be extrapolated to the case of LCDAs for the $B_c$-meson in the limit $m_c \ll m_b$.

Our study shows that the generic parameterization of the leading $B$-meson LCDA in Ref.~\cite{Feldmann:2022uok}, 
together with theoretical constraints from the short-distance behavior computed in this work, and 
estimates for the inverse moments from QCD sum rules in Ref.~\cite{Khodjamirian:2020hob}, provide a 
consistent framework to be used for future phenomenological fits to exclusive $B_s$ decays into energetic hadrons.

\acknowledgments

We thank Danny van Dyk for a critical reading of the manuscript and for helpful comments. This research is supported by the Deutsche Forschungsgemeinschaft (DFG, German Research Foundation) under grant 396021762 --- TRR 257.

\clearpage 

\appendix

\section{Radiative tail of \texorpdfstring{$\tilde \phi_-(\tau)$}{phi-} and general matching calculation}

\label{sec:radiative-tail}

In this appendix, we provide the matching calculation for the OPE of a generic 2-particle light-ray operator in HQET, from which one can also read off the 
result for the radiative tail of the sub-leading 2-particle LCDA $\phi_B^-$ of the $B$ meson.
We define the short-distance expansion of a 2-particle HQET light-ray operator with arbitrary 
Dirac structure $\Gamma$ as
\begin{eqnarray}
{\cal O}_\Gamma(\tau)& = & \bar q(\tau n) \left[\tau n,0\right] \Gamma \, h_v(0) 
	\cr 
 &=& 
  c_{1}^{(3)}(\tau) \, \bar q(0) \, \frac{\slashed n \slashed v}{2} \, \Gamma \, h_v(0)  
  + d_{1}^{(3)}(\tau) \, \bar q(0) \, \frac{\slashed v \slashed n}{2} \, \Gamma \, h_v(0)  
	\cr && {} 
 + c_{1}^{(4)}(\tau) \, \bar q(0) \, (i n \cdot \overleftarrow{D}) \, \frac{\slashed n \slashed v}{2} \, \Gamma \, h_v(0)
 + d_{1}^{(4)}(\tau) \, \bar q(0) \, (i n \cdot \overleftarrow{D}) \, \frac{\slashed v \slashed n}{2} \, \Gamma \, h_v(0)
 \cr && {}
 + c_{2}^{(4)}(\tau) \, \bar q(0) \, (i v \cdot \overleftarrow{D}) \, \frac{\slashed n \slashed v}{2} \, \Gamma \, h_v(0)
 + d_{2}^{(4)}(\tau) \, \bar q(0) \, (i v \cdot \overleftarrow{D}) \, \frac{\slashed v \slashed n}{2} \, \Gamma \, h_v(0)
 \cr && {}
	   + c_{3}^{(4)}(\tau) \, m \, \bar q(0) \, \frac{2\slashed v- \slashed n}{2} \, \Gamma \, h_v(0)
       + d_{3}^{(4)}(\tau) \, m \, \bar q(0) \, \frac{\slashed n}{2} \, \Gamma \, h_v(0)
    + {\cal O}(\tau^2) \,,
    \label{ope2} 
\end{eqnarray}
which for $\Gamma=\slashed n\gamma_5$ reduces to the OPE for ${\cal O}_+(\tau)$ in 
Eq.~(\ref{ope}). Here we find it convenient to separate the individual structures by means of light-cone projectors,
\begin{eqnarray}
 P_+ = \frac{\slashed n \slashed v}{2} = \frac{\slashed n \slashed {\bar n}}{4} \,, \qquad 
 P_- = \frac{\slashed v \slashed n}{2} = \frac{\slashed {\bar n} \slashed n}{4} \, ,
\end{eqnarray}
with $P_\pm^2 = P_\pm$ and $P_+ + P_- =1$.
Taking, on the other hand, the Dirac matrix $\Gamma=\slashed {\bar n}\gamma_5$, we obtain the corresponding OPE which determines the radiative tail of $\phi_B^-(\tau)$,
\begin{eqnarray}
{\cal O}_-(\tau) &=&	\bar q(\tau n) \left[\tau n,0\right] \slashed {\bar n}\gamma_5 \, h_v(0) = \sum_{n=3}^\infty \sum_{k=1}^{K_n} d_k^{(n)}(\tau) \, {\cal O}_k^{'(n)}(0)
\cr 
 &=& d_{1}^{(3)}(\tau) \, \bar q \, \slashed {\bar n}\gamma_5 \, h_v  
\cr && {} + d_{1}^{(4)}(\tau) \, \bar q \, (i n \cdot \overleftarrow{D}) \, \slashed {\bar n}\gamma_5 \, h_v
+ d_{2}^{(4)}(\tau) \, \bar q \, (i v \cdot \overleftarrow{D}) \, \slashed {\bar n} \gamma_5 \, h_v 
\cr 
&& {} + d_{3}^{(4)}(\tau) \,  m \, \bar q \, \slashed v \slashed {\bar n} \gamma_5 \, h_v + \ldots
\end{eqnarray}
The radiative tail of the LCDA $\tilde \phi_B^-(\tau)$ is then given by 
\begin{eqnarray}
	\tilde\phi_-(\tau) =
    d_1^{(3)}(\tau)
    + \bar \Lambda \left(
        \frac23 \, d_1^{(4)}(\tau)
        + d_2^{(4)}(\tau)
    \right)
	- m \left(
         d_3^{(4)}(\tau)
         - \frac13 \, d_1^{(4)}(\tau)
     \right)
     + {\cal O}(\tau^2) \,.
 \label{phiexp2}
\end{eqnarray}
Here, we have used again the hadronic matrix elements of the local HQET
operators,
\begin{eqnarray}
	\langle 0| {\cal O}_1^{'(3)} | \bar{B}(v)\rangle 
	= i \, m_B f_B^{\rm HQET}  \,,
\end{eqnarray}
and
\begin{eqnarray}
&&	\frac{\langle 0| {\cal O}_1^{'(4)} |\bar{B}(v)\rangle}{\langle 0| {\cal O}_1^{'(3)} | \bar{B}(v)\rangle} = \frac{2\bar\Lambda + m}{3} \,, 
	\quad 
	\frac{\langle 0| {\cal O}_2^{'(4)} |\bar{B}(v)\rangle}{\langle 0| {\cal O}_1^{'(3)} | \bar{B}(v)\rangle} = \bar\Lambda \,,
	\quad
	\frac{\langle 0| {\cal O}_3^{'(4)} |\bar{B}(v)\rangle}{\langle 0| {\cal O}_1^{'(3)} | \bar{B}(v)\rangle} = -  m \,.
\end{eqnarray} 

According to our detailed discussion in the main text, the contribution of the vertex diagram (a) to the matching coefficients can then be obtained from a Fourier integral with subtractions, 
\begin{eqnarray}
    \int_0^\infty d\omega \left( e^{-i\omega \tau}  - 1 + i \omega\tau +\ldots \right) I_a^\Gamma(\omega,m,k) \,,
\end{eqnarray}
where in the integrand the function $I_a^\Gamma(\omega,m,k)$ is now calculated for general external on-shell quark states, where the light quark carries momentum $k^\mu$: 
\begin{eqnarray}
	I^\Gamma_a(\omega, m, k) &=&
	-i \int[d\ell] \, \delta(\omega - n\cdot (k - \ell)) \;
	\frac{
		\bar v(k) \, \slashed{v}
		(- \slashed k + \slashed \ell + m)
		\, \Gamma \, u(v)
	}{
		[(k-\ell)^2 - m^2 + i0][v\cdot \ell + i0][\ell^2 + i0] 
	} \,. 
 \cr &&
\end{eqnarray}
Expanding in the external momentum and mass, we find
\begin{eqnarray}
    I_a^\Gamma(\omega,m,k) &=& 
    \frac{\Gamma(1+\epsilon)}{\omega}
    \left( \frac{\mu^2 e^{\gamma_E}}{\omega^2} \right)^\epsilon 
    \cr && {} \times 
    \bar v(k) \, \left\{ 
     \left( 2 + (1+2\epsilon) \, \frac{n\cdot k}{\omega} 
        + (4+2\epsilon) \, \frac{v\cdot k}{\omega} \right) 
        \frac{\slashed n \slashed v}{2}  \right.
        \cr && \left. \qquad {} 
        + \left( -2 - (1+2\epsilon) \, \frac{n\cdot k}{\omega} 
        + (2-2\epsilon) \, \frac{v\cdot k}{\omega} \right)   \frac{\slashed v \slashed n}{2} \right. 
        \cr && \left. \qquad {}
        - \frac{m}{\omega} \, \slashed v + {\cal O}(\omega^{-2})
    \right\} \Gamma \,  u(v) \, .
\end{eqnarray}
Performing the $\omega$-integration, we end up with
\begin{eqnarray}
&& \int_0^\infty d\omega \left( e^{-i\omega \tau}  - 1 + i \omega\tau +\ldots \right) I_a^\Gamma(\omega,m,k)
\cr 
  &=& 
    \bar v(k) \, \left\{ 
     \left( -\frac{1}{\epsilon} - 2L  + \left(\frac{1}{2\epsilon} + L \right) i \tau (n\cdot k)
        + \left(\frac{2}{\epsilon} + 4 L - 3 \right) i\tau (v\cdot k) \right) 
        \frac{\slashed n \slashed v}{2}  \right.
        \cr && \left. \qquad \  {} 
        + \left( \frac{1}{\epsilon} + 2 L  - \left( \frac{1}{2\epsilon} + L \right) i\tau (n\cdot k) 
        + \left( \frac{1}{\epsilon} + 2 L -3 \right) i\tau(v\cdot k) \right) \frac{\slashed v \slashed n}{2}\right. 
        \cr && \left. \qquad {}
        - \left( \frac{1}{2\epsilon} + L - 1 \right) i\tau m \, \slashed v + {\cal O}(\tau^{2})
    \right\} \Gamma \,  u(v) \,.
\end{eqnarray}
From this we can easily read off the contribution of the vertex diagram (a) to the individual Wilson coefficients.
The contribution from diagram (b) (Wilson-line coupled to heavy quark) can be derived in a similar manner, with
\begin{eqnarray} 
&& \int_0^\infty \, d\omega \left(e^{i\omega \tau} - 1 + i \omega\tau  \right) I_b^\Gamma(\omega,k) 
\cr 
  &=&  \left(1 -i\tau (n \cdot k) + {\cal O}(\tau^2) \right)
	\left( -\frac{1}{\epsilon^2}
	- \frac{2L}{\epsilon}
	- 2 L^2 - \frac{5\pi^2}{12} \right) 
	\bar v(k) \, \frac{ \slashed n \slashed v + \slashed v \slashed n}{2} \, \Gamma \, u(v) \,. 
\end{eqnarray}
Finally, as discussed in the main part of the text, the diagram (c) with the Wilson line coupled to the light quark does not contribute to the matching. With this we confirm our results for the matching coefficients in the expansion of 
${\cal O}_+(\tau)$.
For the matching coefficients relevant to the LCDA $\phi_B^-$ we obtain after $\overline{\mathrm{MS}}$ renormalization
\begin{eqnarray}
	d_1^{(3)}(\tau) &=& 1 -\frac{\alpha_s C_F}{4\pi} 
	\left( 2 L^2-2 L + \frac{5\pi^2}{12}\right) + {\cal O}(\alpha_s^2) \,, \cr 
	d_1^{(4)}(\tau) &=& -i\tau \left[ 1 - \frac{\alpha_s C_F}{4\pi} \left( 2 L^2-L + \frac{5\pi^2}{12} \right)  + {\cal O}(\alpha_s^2) \right]  \,, \cr 
	d_{2}^{(4)}(\tau) &=& -i\tau \left[ - \frac{\alpha_s C_F}{4\pi} \left( 2L-3 \right) + {\cal O}(\alpha_s^2)\right]  \,, \cr 
    d_{3}^{(4)}(\tau) &=& -i \tau \left[ \frac{\alpha_s C_F}{4\pi} \left( L -1 \right) + {\cal O}(\alpha_s^2) \right] \,.
    \label{eq:d34minus}
\end{eqnarray}
Plugging this result into the expression for the LCDA $\tilde\phi_B^-(\tau)$,
we obtain
\begin{eqnarray}
    \tilde\phi_B^-(\tau) &=& 
    \left[1- i \tau \, \frac{2 \bar\Lambda  + m }{3} \right]\left[  1 - \frac{\alpha_s C_F}{4\pi} \left( 2 L^2 - 2 L + \frac{5\pi^2}{12} \right) \right]
    \cr 
    && {} + i \tau \bar\Lambda \, \frac{\alpha_s C_F}{4\pi} \left( \frac83  \, L - 3
 \right) 
      + i \tau m \, \frac{\alpha_s C_F}{4\pi} \left( \frac{4}{3}\, L - 1 \right)  + {\cal O}(\alpha_s^2,\tau^2) \,.
    \label{phitauexp2}
\end{eqnarray}
We note that the terms in the second line are identical for $\tilde\phi_B^-(\tau)$ and $\tilde\phi_B^+(\tau)$.

\section{Numerical results for the LCDA of the \texorpdfstring{$B_c$}{Bc} meson}

\label{sec:Bc}

We briefly give the analogous discussion to Sec.~\ref{sec:num} for the case of the $B_c$.
Due to the higher mass of the charm quark, a larger renormalization scale is needed.
We consider here $\mu_0 = 2~\mathrm{GeV}$ such that
$\alpha_s(\mu_0) \simeq 0.3$,
and $\omega_0 = 1.18~\mathrm{GeV}$ (for $x_0 = 1$ and $n_0 = 1/3$).
We take $M_{B_c} = 6274.47\pm0.32~\mathrm{MeV}$ and $m_c(m_c) = 1.27\pm0.2~\mathrm{GeV}$ \cite{ParticleDataGroup:2022pth},
from which we calculate $m_c(\mu_0) \simeq 1.10~\mathrm{GeV}$ using the software \texttt{RunDec}~\cite{Herren:2017osy}.
This corresponds to
\begin{eqnarray}
    \bar\Lambda^{(c)}_a(\mu_0, x_0 = 1) \simeq 1244~\mathrm{MeV} \,,\quad
    m^{(c)}_a(\mu_0, x_0 = 1) \simeq 995~\mathrm{MeV} \,,\quad
\end{eqnarray}
which yields the coefficients
\begin{eqnarray}
    a^{(c)}_0 &\simeq&
          1.87
        - 0.77 \, \frac{\bar\Lambda^{(c)}_a}{1244~\mathrm{MeV}}
        + 0.12 \, \frac{m^{(c)}_a}{995~\mathrm{MeV}}
        + a^{(c)}_2 = 1.22 + a_2^{(c)} \,,
        \cr
    a^{(c)}_1 &\simeq & 
        0.93
        - 0.71 \, \frac{\bar\Lambda^{(c)}_a}{1244~\mathrm{MeV}}
        + 0.13 \, \frac{m^{(c)}_a}{995~\mathrm{MeV}}
        + 2 a^{(c)}_2 = 0.35 + 2 \, a_2^{(c)}
        \,.
\end{eqnarray}
The convergence criterion then leads to
\begin{eqnarray}
    -0.32 < a^{(c)}_2 < 0.25 \,,
\end{eqnarray}
and the inverse moment reads
\begin{eqnarray}
    \lambda_{B_c} &\simeq&\frac{885~\mathrm{MeV}}
    {
        1.40
        -0.57 \frac{\bar\Lambda^{(c)}_a}{1244~\mathrm{MeV}}
        +0.090 \frac{m^{(c)}_a}{995~\mathrm{MeV}}
        + a^{(c)}_2
    } 
    = \frac{885~\mathrm{MeV}}
    {
        0.91 + a^{(c)}_2
    } \,.
\end{eqnarray}

\begin{figure}[t!p]
\begin{center}
	\begin{tabular}{p{0.4\textwidth} p{0.4\textwidth}}
	  \vspace{0pt}\includegraphics[scale=0.8]{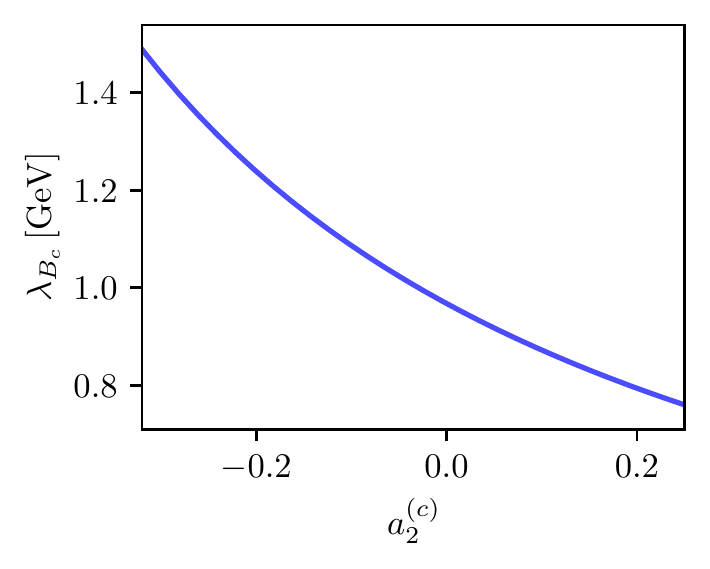} &
	  \vspace{0pt}\includegraphics[scale=0.8]{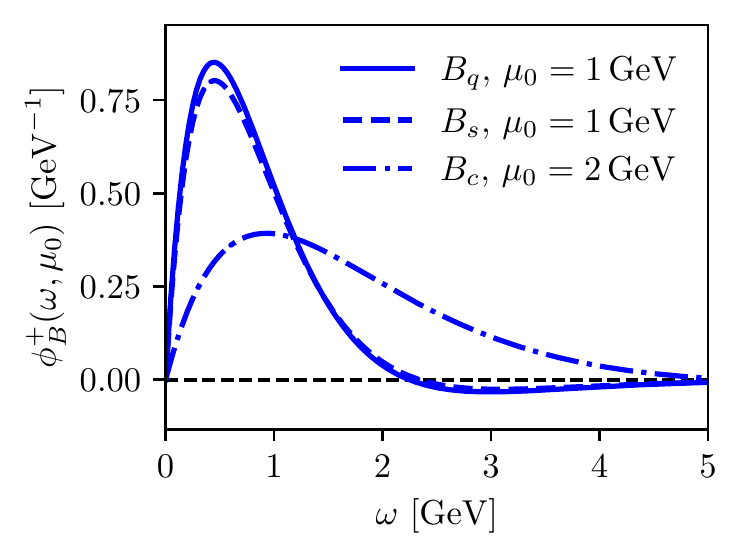}
	\end{tabular}
\end{center}
\caption{\label{fig:numerical-results-B-c}
     Left: estimate for the inverse moment $\lambda_{B_c}$ of the $B_c$-meson LCDA  as a function of the expansion coefficient $a_2$.
     Here the blue solid line corresponds to the central value following from the 
     constraint of the dimension-3 and dimension-4 operators on the radiative tail, based on the truncated parameterization.
     Right: comparison of the central values for our estimates of the $B_q$, $B_s$ and $B_c$ LCDAs. (Here the parameters $a_2^{(q,s,c)}$ 
     are taken to be zero for simplicity.)
     }
\end{figure}

Fig.~\ref{fig:numerical-results-B-c} shows the inverse moment as a function of $a^{(c)}_2$.
Interestingly, the expansion parameters $a_k^{(c)}$ turn out to be not very different from the light-quark case.
However, one should be aware that both the renormalization scale $\mu_0$ and the reference momentum $\omega_0$ 
are strikingly different for the two cases. As a consequence, the inverse moment $\lambda_{B_c}$ comes out significantly
larger than for light spectator quarks, close to the non-relativistic limit (for which $m_c \approx \bar\Lambda_{B_c} \approx \lambda_{B_c}$).
For comparison, we have collected the relevant quantities for $B_q$, $B_s$, and $B_c$ mesons in Tab.~\ref{tab:comp-q-s-c}.
In addition, we have plotted the central values of our estimates for the LCDAs of $B_q$, $B_s$ and $B_c$ mesons in momentum space
on the r.h.s.\ of Fig.~\ref{fig:numerical-results-B-c}. Here, for illustrative purposes, 
the expansion parameters $a_2$ have been set to zero for simplicity. As one would have intuitively expected, the effect of increasing the spectator mass is 
to lower the value and to increase the width of the peak at small light-cone momentum. On the other hand, if one rescaled the LCDAs by the corresponding 
estimate for the inverse moment, the three curves for the quantity $\lambda_B \, \phi_B^+(\omega/\lambda_B)$ would look almost indistinguishable.

\begin{table}[t!p]
\begin{center}
\begin{tabular}{|c || c | c || c |}
  \hline 
    & $B_q$ & $B_s$ & $B_c$ \\
    \hline \hline
    $\mu_0$  & 1~GeV & 1~GeV & 2~GeV \\
    $\omega_0$ &  594~MeV & 594~MeV &  1.18~GeV \\
    \hline 
    $\bar \Lambda_a$ & 367~MeV & 437~MeV & 1.24~GeV \\
    $ m_a$ & 0 & 106~MeV & 1.00~GeV \\
    \hline 
    $a_0-a_2$ & 1.31 & 1.24 &  1.22 
    \\
    $a_1-2a_2$ & 0.47 & 0.42 & 0.35
    \\
     \hline
    $\lambda_B$ & $(380, 690)$~MeV & $(390, 730)$~MeV & $(0.76, 1.49)$~GeV
    \\ \hline
  \end{tabular}
\end{center}
\caption{\label{tab:comp-q-s-c} Comparison between key quantities characterizing the LCDAs of $B_q$, $B_s$, and $B_c$ mesons (central values only).}
\end{table}

\section{Possible impact of dimension-5 operators}
\label{sec:app-dim-5}

In this appendix we study the possible impact of dimension-5 operators on the determination of the expansion coefficients $a_k$.
At tree level, the following dimension-5 local operators have non-vanishing Wilson coefficients \cite{Grozin:1996pq}
$$
  \bar q \, iG^{\mu\nu} \, h_v \,, \qquad \frac12 \, \bar q \{ i\overleftarrow{D}^\mu, i \overleftarrow{D}^\nu \} \, h_v \,.
$$
The $B$-meson matrix element of the first operator does not have an explicit mass dependence. 
Translating the definition from Ref.~\cite{Grozin:1996pq} into the covariant trace formalism, we have
\begin{align}
    \frac{\langle 0|\bar q_\beta \, iG^{\mu\nu} \, (h_v)_\alpha| \bar{B}(v) \rangle}{\langle 0| {\cal O}_1^{(3)}| \bar{B}(v)\rangle }
    &= \frac14 \left[(1+\slashed v) \left( \frac{\lambda_H^2-\lambda_E^2}{3} \, (\gamma^\mu v^\nu -\gamma^\nu v^\mu) - \frac{\lambda_H^2}{3} \,
    i\sigma^{\mu\nu}\right) \gamma_5 \right]_{\alpha\beta} \,.
\end{align}
The mass dependence for the second operator can be obtained by taking into account the quark mass in the Dirac equation for the light quark.
In this way we find
\begin{align}
    & \frac{\frac12 \, \langle 0|\bar q_\beta \, \{ i \overleftarrow{D}^\mu,i \overleftarrow{D}^\nu \} \, (h_v)_\alpha| \bar{B} (v)\rangle}{\langle 0| {\cal O}_1^{(3)}| \bar{B}(v)\rangle}
    \cr 
    = & - \frac14 \left[(1+\slashed v) \left( 
      \frac{6\bar\Lambda^2 + 2 \lambda_E^2 + \lambda_H^2 - 2 m \bar\Lambda - m^2}{3} \,  v^\mu v^\nu \right. \right. 
      \cr 
      & \qquad  
     \left. \left. - \frac{\bar \Lambda^2 +\lambda_E^2+\lambda_H^2-m^2}{3} \, g^{\mu\nu} 
     - \frac{2\bar\Lambda^2 + \lambda_E^2- 2 m \bar\Lambda}{6} \, (\gamma^\mu v^\nu +\gamma^\nu v^\mu )
     \right) \gamma_5 \right]_{\alpha\beta} \,.
\end{align}
With this the \emph{tree level} expressions for the second Mellin moment of the momentum-space $B$-meson LCDAs
follows as
\begin{align} 
    \langle \omega^2 \rangle_+ 
        \equiv  \int_0^\infty d\omega \, \omega^2 \, \phi_B^+(\omega) \Big|_{\rm tree} 
        &= \frac{\bra{0} \bar q \, (i n\cdot \overleftarrow{D})^2 \, \slashed n \gamma_5 h_v \ket{\bar{B}(v)}}{\langle 0| {\cal O}_1^{(3)}| \bar{B}(v)\rangle } \cr
        &= \frac{6\bar\Lambda^2+2\lambda_E^2+\lambda_H^2 - 2 m\bar\Lambda - m^2}{3} \,, 
        \cr 
    \langle \omega^2 \rangle_-  \equiv \int_0^\infty d\omega \, \omega^2 \, \phi_B^-(\omega) \Big|_{\rm tree}
        &= \frac{\bra{0} \bar q \, (i n\cdot \overleftarrow{D})^2 \, \slashed{\bar n} \gamma_5 h_v \ket{\bar{B}(v)}}{\langle 0| {{\cal O}}_1^{'(3)}| \bar{B}(v)\rangle } \cr
        &= \frac{2\bar\Lambda^2+\lambda_H^2 + 2 m\bar\Lambda - m^2}{3} \,, 
\end{align}
generalizing the findings in Ref.~\cite{Grozin:1996pq} to the massive case.
In the non-relativistic limit, $\bar\Lambda \to m$, $\lambda_{E,H}^2 \to 0$, this consistently reduces to
$\langle \omega^2\rangle_\pm = m^2$.

Including the numerically dominant 1-loop corrections at dimension-5 level, making use of the results in Eq.~(7) in Ref.~\cite{Kawamura:2008vq}, leads to
\begin{eqnarray}
    \tilde\phi_+(\tau) &=& 
    \left[1- i \tau \, \langle\omega\rangle_+  
    - \tau^2 \, \frac{\langle \omega^2\rangle_+}{2} \right]\left[  1 - \frac{\alpha_s C_F}{4\pi} \left( 2 L^2+2 L + \frac{5\pi^2}{12} \right) \right]
    \cr 
    && {} + i \tau \bar\Lambda \, \frac{\alpha_s C_F}{4\pi} \left( \frac83 \, L -3 \right) 
      + i \tau m \, \frac{\alpha_s C_F}{4\pi} \left( \frac43 \, L - 1 \right) 
      \cr 
    && {} +  \tau^2 \bar{\Lambda}^2
        \frac{\alpha_s C_F}{4\pi} 
        \left( \frac{10}{3}L-\frac{35}{9} 
         + {\cal O}\left(\frac{m}{\bar\Lambda} \right) + {\cal O}\left( \frac{\lambda_{E,H}^2}{\bar\Lambda^2} \right) \right) \cr
    && {}   + {\cal O}(\alpha_s^2) + {\cal O}(\tau^3) \,.
    \label{phitauexpdim5}
\end{eqnarray}
Here, in the first line we took into account the universal double-logarithmic corrections proportional to $\langle \omega^2\rangle_+$, while for the single-logarithmic 
corrections to the dimension-5 contributions we only included the terms proportional to $\bar\Lambda^2$. The missing $\alpha_s$ corrections at dimension-5 level are numerically 
suppressed by the small values for light quark masses and $\lambda_{E,H}^2$, as indicated.

\begin{table}[t!p]
\begin{center} \small 
	\begin{tabular}{c | ccc | ccc || c |}
		\cline{2-8}
 & \multicolumn{3}{c|}{tree-level, pole-scheme} & \multicolumn{3}{c||}{1-loop$^{(*)}$, pole-scheme} & 1-loop, $a$-scheme 
 \\
 $n_0=1/3$ & dim-3 & dim-4 & dim-5 & dim-3 & dim-4 & dim-5 & dim-4
 \\
 & $K=0$ & $K=1$ & $K=2$ & $K=0$ & $K=1$ & $K=2$ & $K=2$
 \\
 \hline \hline 
 \multicolumn{1}{|c|}{$a_0^{(q)}$} & $1$ & $1.44$ & $1.54$ & $0.78$ & $1.07$ & $1.07$ & $(0.98, 1.51)$
 \\
\multicolumn{1}{|c|}{$a_1^{(q)}$} & -- & $0.44$ & $0.65$ & -- & $0.26$ & $0.23$ & $(-0.19, 0.87)$
 \\
 \multicolumn{1}{|c|}{$a_2^{(q)}$} & -- & -- & $0.11$ & -- & -- & $-0.03$ & $(-0.33, 0.20)$
 \\
 \hline 
 \multicolumn{1}{|c|}{$a_0^{(s)}$ }& $1$ & $1.37$ & $1.43$  &  $0.78$ &  $0.99$ &  $0.96$ & $(0.92, 1.45)$
 \\ 
\multicolumn{1}{|c|}{ $a_1^{(s)}$ }&	-- & $0.37$ & $0.49$ & -- & $0.20$ & $0.10$ & $(-0.22, 0.84)$
 \\
\multicolumn{1}{|c|}{ $a_2^{(s)}$}& -- & -- & $0.06$ & -- & -- & $-0.06$ & $(-0.32, 0.21)$
 \\
 \hline
	\end{tabular}
\end{center}
\caption{\label{tab:comp-dim-5} 
	Comparison of different estimates for the expansion parameters, depending on the order of perturbation theory,
 the order of the OPE, and the considered truncation level $K$. The last column refers to the default case discussed in the main body of the text. ${}^{(*)}$ Notice that 
 part of the 1-loop corrections to the dimension-5 operators have been neglected, see the discussion around Eq.~(\ref{phitauexpdim5}).}
\end{table}

Adjusting the parameters of the generic parameterization (\ref{phitau_ak}) for truncation level $K=2$ and using the additional information on the second moment, 
the tree-level expressions for the expansion coefficients then follow as
\begin{eqnarray}
	a_0 &=&  1  + {}\left(1- \frac{\langle \omega \rangle_+}{2\omega_0}\right) 
     + {} \frac34 \left( 1 -\frac{\langle \omega_+\rangle}{\omega_0}
     + \frac{\langle \omega^2\rangle_+}{6\omega_0^2} \right) + \ldots  \,, \cr 
	a_1 &=&  \phantom{1  + {}}  \left(1- \frac{\langle \omega\rangle_+}{2\omega_0}\right) + \frac32 \left( 1 -\frac{\langle \omega_+\rangle}{\omega_0}
     + {} \frac{\langle \omega^2\rangle_+}{6\omega_0^2} \right) + \ldots \,, \cr 
    a_2 &=& \phantom{1  + {}\left(1- \frac{\langle \omega\rangle_+}{2\omega_0}\right) + {}} 
     {} \frac34 \left( 1 -\frac{\langle \omega_+\rangle}{\omega_0}
     + {} \frac{\langle \omega^2\rangle_+}{6\omega_0^2} \right) + \ldots \,,
\end{eqnarray}
where the first Mellin moment is given by $\langle \omega\rangle_+= (4\bar\Lambda- m)/3$, see above.
The numerical results are collected in the first three columns of Table~\ref{tab:comp-dim-5}, where we also give the results for lower truncation level ($K=0,1$) 
which are obtained by dropping the corresponding columns in the formula above. From this we already get a rough estimate of the numerical convergence of our approach.
Here, for the HQET parameters we now consider the pole-mass scheme, with the central value for the 
$b$-quark pole mass taken as $m_b \simeq 4.78$~GeV from Ref.~\cite{ParticleDataGroup:2022pth}, which corresponds to
$$
\bar\Lambda^{(q)}  \simeq 500~{\rm MeV} \,, 
\qquad 
\bar\Lambda^{(s)}  \simeq 590~{\rm MeV} \,,
$$
together with $\lambda_E^2=0.01$~GeV$^2$ and $\lambda_H^2=0.15$~GeV$^2$ from Ref.~\cite{Rahimi:2020zzo}.
Notice that the corresponding 1-loop expressions for $\Lambda_a$ would come out significantly larger than the values considered in 
Section~\ref{sec:num} (which are obtained from a 1-loop comparison with the DA-scheme defined in \cite{Lee:2005gza}). 
The difference between the two treatments of the HQET mass parameters may thus give some handle to estimate 
the associated scheme dependence of our results.
In the same spirit, we now use the $\overline{\rm MS}$ mass values for the light quarks (see above).
For comparison, in the last column of Table~\ref{tab:comp-dim-5} we show the estimated ranges for the expansion coefficients as obtained from Section~\ref{sec:num}, where the parameter $a_2$
is varied within the interval fixed by the ad-hoc convergence criterion, and the mass parameters are taken in the $a$-scheme. We observe that for the considered hadronic input values, the resulting central values for the expansion coefficients are compatible with the intervals obtained in Section~\ref{sec:num}.

Based on Eq.~(\ref{phitauexpdim5}) we also determined the 1-loop result for the expansion coefficients at truncation level $K=0,1,2$, whose central numerical values are again collected in Table~\ref{tab:comp-dim-5}. We observe that the 1-loop corrections somewhat improve the convergence of the procedure compared to the tree-level case, while the 
central values for $a_{0,1,2}$ are entirely consistent with the ranges obtained in Section~\ref{sec:num}.
The central values also confirm the trend of decreasing values of $a_{1,2}$ for increasing spectator quark masses, 
as inferred from the comparison with the QCD sum rule results for $\lambda_{B_q}$ and $\lambda_{B_s}$
in Section~\ref{sec:num}.
Altogether, this further supports our conclusion that information from the OPE can be consistently implemented in global phenomenological analyses 
in the framework of QCD factorization, based on the parameterization proposed in Ref.~\cite{Feldmann:2022uok}.

\bibliographystyle{JHEP-2}
\bibliography{bib}

\end{document}